\documentclass[reprint,showpacs,aps,prl,amsmath,amssymb,floatfix, twocolumn]{revtex4-1}

\usepackage{graphicx,epstopdf}% Include figure files
\usepackage{bm}% bold math
\usepackage{mathrsfs}
\usepackage{color}
\usepackage[bookmarks,colorlinks,linkcolor=blue,anchorcolor=black,urlcolor=blue,citecolor=blue]{hyperref}
\usepackage{natbib}
\usepackage{CJK}
\allowdisplaybreaks[3]

\begin{document}
\begin{CJK*}{UTF8}{gbsn}

\title{Beam focusing and reduction of quantum uncertainty in width at the few-photon level via multi-spatial-mode squeezing}

\author{Lida Zhang~(\CJKfamily{gbsn}张理达)$^{1}$}
\author{G. S. Agarwal$^{1,2}$}
\author{M. O. Scully$^{1,3,4}$}
\affiliation{$^{1}$ Texas A \& M University, College Station, Texas}
\affiliation{$^{2}$ Department of Physics and Astronomy, Texas A \& M University, College Station, Texas 77843}
\affiliation{$^{3}$ Baylor University, Waco, TX 76798}
\affiliation{$^{4}$ Princeton University, Princeton, NJ 08544}
% \affiliation{$^{1}$ The Institute for Quantum Science and Engineering, Texas A \& M University, College Station, Texas}
% \affiliation{$^{3}$ Department of Physics, Baylor University, Waco, TX 76798}
% \affiliation{$^{4}$ Department of Mechanical and Aerospace Engineering, Princeton University, Princeton, NJ 08544}
\date{\today}

\begin{abstract}
We show for the first time that it is possible to realize laser beam focusing at the few-photon level in the four-wave-mixing process, and at the same time, reducing the quantum uncertainty in width. The reduction in quantum uncertainty results directly from the strong suppression of local intensity fluctuations. This surprising effect of simultaneous focusing and reduction of width uncertainty is enabled by multi-spatial-mode squeezing, and is not possible via any classical optical approach or single-spatial-mode squeezing. Our results open promising possibilities for quantum-enhanced imaging, bio-sensing and metrology including measurements of nanometer displacement. 
\end{abstract}

% \pacs{11.30.Er, 42.25.Bs, 42.50.Gy, 03.65.Ca} 

% PACS numbers: 03.65.Xp, 03.65.Ca, 03.67.Ac, 03.67.Hk

% 03.65.Xp 	Tunneling, traversal time, quantum Zeno dynamics
% 03.65.Ca  Formalism
% 03.67.Ac  Quantum algorithms, protocols, and simulations
% 03.67.Hk   Quantum communication

%11.30.Er Charge conjugation, parity, time reversal, and other discrete symmetries
%03.65.Ud 03.65.Ud 	Entanglement and quantum nonlocality 

% 42.25.Bs Wave propagation, transmission and absorption [
% 42.82.Et 	Waveguides, couplers, and arrays (

% 42.50.Gy  	Effects of atomic coherence on propagation, absorption, and amplification of light;  
% 42.65.An Optical susceptibility, hyperpolarizability 

\maketitle

\end{CJK*}

% \section{Introduction}

{\it Introduction.}
% From a quantum perspective, a laser beam is always accompanied with beam width uncertainty due to inevitable intensity fluctuations. 
% Focusing a laser beam to a small spot is of great practical and fundamental importance for a variety of applications. 
It is well known that a laser beam can be focused via classical linear or nonlinear optical techniques, e.g., lens or self-Kerr effects~\cite{Boyd2008NO}. As the laser intensity decreases to the few-photon level, focusing can only be possible via linear optics since nonlinear effects become negligible. 
However, at the few-photon level, 
% When the laser beam is attenuated to a few-photon level, 
the quantum intensity fluctuations which leads to the laser beam width uncertainty becomes critically important. The uncertainty scales as $w_{0}/\sqrt{N}$ with $w_{0}$ and $N$ being the beam width and photon numbers respectively~\cite{Treps2003Science,Chille2015OE}, i.e., the weaker the laser beam becomes, the stronger the width uncertainty will be. Obviously, classical approaches based on linear optics can be only used to focus the beam but is unable to reduce the width uncertainty. 
It thus significantly limits its applications in quantum-based precision measurements where only very-weak light at the few-photon level is allowed in order to avoid damages, for example, in biological systems~\cite{Taylor2013NPhoton,Taylor2014PRX}. 

In recent years, multi-spatial-mode (MSM) quadrature squeezing, which explores the transverse spatial degree freedom of light, has received intensive investigations due to its promising applications in a variety of directions such as quantum entanglement and information~\cite{Gatti1999PRL,Treps2005PRA,Lassen2007PRL,Boyer2008Science,Marino2009Nature,Gatti2009PRL,Janousek2009NPhoton,Tasca2011PRA,Kang2012PRL,Edgar2012NComm,Pooser2014PRA,Kovlakov2017PRL}, ultra-sensitive measurement of nanometer displacement~\cite{Treps2004JOB,Pooser2015Optica}, quantum-enhanced metrology like sub-shot-noise or super-resolution quantum imaging imaging~\cite{Kolobov2000PRL,Treps2002PRL,Brambilla2008PRA,Lopaeva2013PRL} and detection of gravitational waves~\cite{LIGO2013NPhoton,Oelker2014OE,Toyra2017PRD}. 
% In particular, MSM squeezing has proven  to offer great capabilities in achieving super-resolution imaging, measurements of nanometer displacement, sub-shot-noise quantum imaging in biological systems. 
In general, MSM squeezing involves a large number of squeezed spatial modes, implying localized spatial squeezing and thus reduction of local intensity fluctuation. A series of elegant experiments have been performed to demonstrate MSM squeezing in atomic system~\cite{Boyer2008PRL,Corzo2011OE,Chalopin2011OE,Gabriel2011PRL,Corzo2012PRL,Embrey2015PhysRevX,Du2017APL}. It is straightforward to envision that MSM squeezing would result in stronger suppression of local intensity fluctuations in the transverse plane~\cite{Lugiato1997JOSAB} and thus reduction of beam width uncertainty. 

% One may thus be able to reduce the width uncertainty using MSM squeezing.

Here we demonstrate a surprising effect enabled by MSM squeezing, which is, focusing of a very weak laser beam at the few-photon level and simultaneously remarkable reduction of beam width uncertainty due to the strong suppression of local intensity fluctuation. Our scheme is specifically explained in a conjugate four-wave mixing process in atomic gases, but should be also possible in other systems. 

\begin{figure}[b!]
 \centering
 \includegraphics[width=0.3\textwidth]{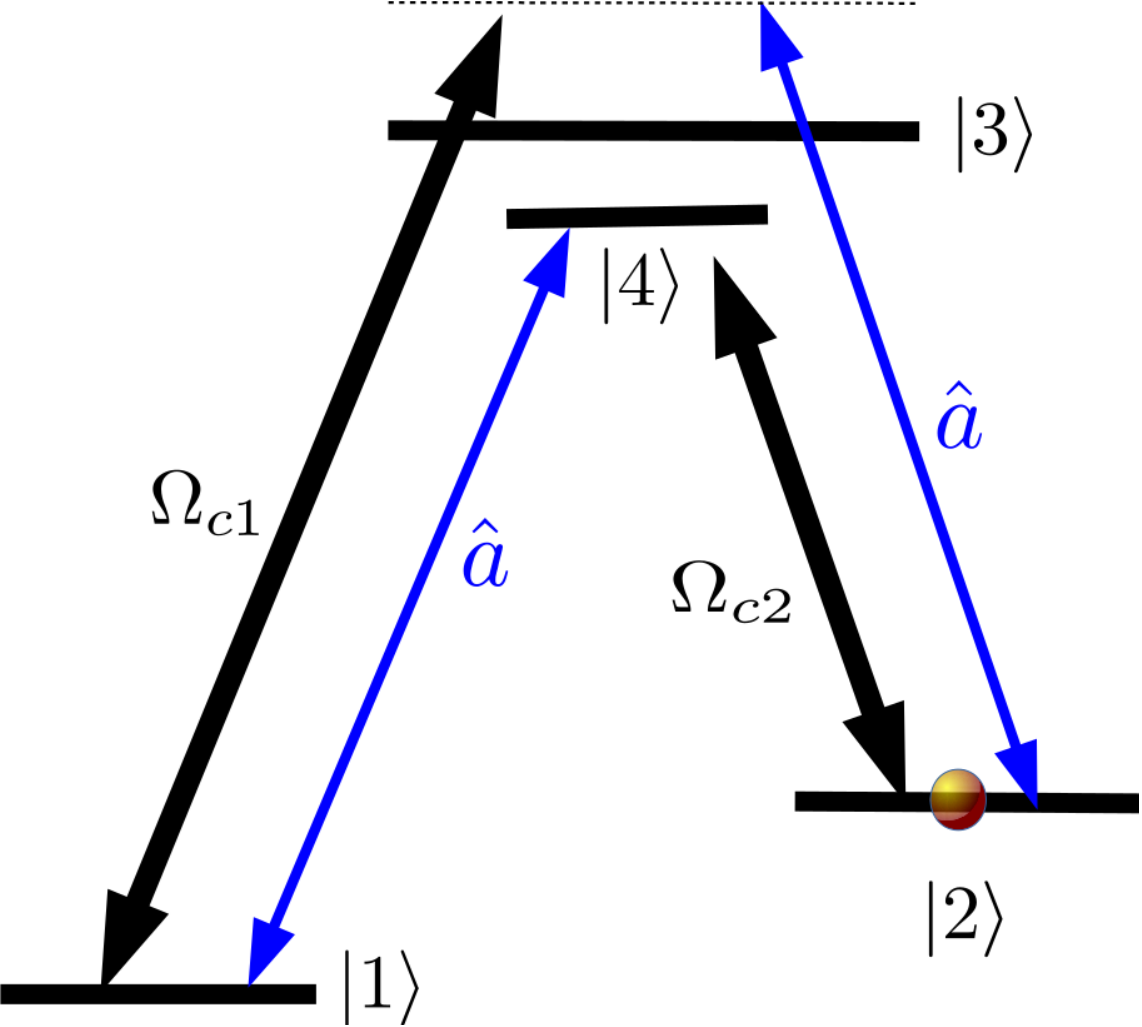} 
 \caption{(Color online) Conjugate FWM process to realize beam focusing and reduction of beam width uncertainty, where two classical control fields $\Omega_{c1}$ and $\Omega_{c2}$, and a quantum probe $\hat{a}_{p}$ are applied. Experimentally, the four levels can be found from, for example, the D1 line of $^{87}$Rb atoms.}
 \label{ConjugateFWM}
\end{figure}

{\it Theoretical model.}
As shown in Fig.~\ref{ConjugateFWM}, our scheme to realize beam focusing is implemented in a conjugate four-wave mixing process~\cite{Lukin1999PRL,Lu1998OL,McCormick2007OL,Jasperse2011OE,Boyer2008PRL,Boyer2008Science,Marino2009Nature,Embrey2015PhysRevX,Zhang2015SR,Du2017APL}.
The interacting Hamiltonian can be written as
\begin{align}
 \hat{H}=&-\frac{\hbar\mathcal{N} }{L}\int^{L}_{0}dz[\Delta\hat{\sigma}_{22} +\Delta_{c1}\hat{\sigma}_{33} +(\Delta+\Delta_{c2})\hat{\sigma}_{44}   \nonumber\\[2ex]
%  &+\Delta_{c1}\hat{\sigma}_{33} +(\Delta_{c1}+\Delta_{c2}-\Delta_{p1})\hat{\sigma}_{44} \nonumber\\[3ex]
 &+\Omega_{c1}\hat{\sigma}_{31} + \Omega_{c2}\hat{\sigma}_{42} + g_{32}\hat{a}\,\hat{\sigma}_{32} + g_{41}\hat{a}\,\hat{\sigma}_{41} + h.c.]\,,
\end{align}
where $\mathcal{N}$ is the number of atoms in the quantization volume of length $L$. $g_{j}$ is the coupling coefficient for the quantum probe $\hat{a}$ which is defined as $g_{j}=\mu_{j}\mathcal{E}_{p}/\hbar$ with $\mu_{j}$ being the dipole moment of the corresponding transition and $\mathcal{E}_{p}=\sqrt{\hbar\omega_{p}/(2\epsilon_0 V)}$ the electric field of a single probe photon ($j\in\{32,41\}$). Here for simplicity $g_{j}$ is assumed to be real and also we will set $g_{32}=g_{41}=g$ in the following. Here we have further simplified the collective atomic operators $\hat{\sigma}_{ij}(\bm{r},t)=\sum_{n}|i_{n}(t)\rangle\langle j_{n}(t)|\delta(\bm{r}-\bm{r}_{n})$ as $\hat{\sigma}_{ij}$ with $\delta(\bm{r}-\bm{r}_{n})$ being the Dirac delta function. $\Omega_{cj}$ ($j\in\{1,2\}$) are the Rabi frequencies of the classical control fields respectively.
$\Delta_{c1}=\omega_{c1}-(\omega_{3}-\omega_{1}),\Delta_{c2}=\omega_{c2}-(\omega_{4}-\omega_{2})$ and $\Delta_{p}=\omega_{p}-(\omega_{3}-\omega_{2})$ are the detunings for the corresponding fields, and $\Delta=\Delta_{c1}-\Delta_{p}$ is the two-photon detuning between $\Omega_{c1}$ and $\hat{a}$, here we assume $2\omega_{p}=\omega_{c1}+\omega_{c2}$, leading to $\Delta_{p}=(\Delta_{c1}+\Delta_{c2}+\omega_{21}+\omega_{43})/2$ where $\omega_{ij}=\omega_{i}-\omega_{j}$. In the Hamiltonian, we have also assumed the two control fields are much stronger than the two quantum fields such that they can be considered as classical. Furthermore, since the control fields are chosen to be far-detuned from the atomic transition, their propagation in the medium would be the same as in vacuum where only free-space diffraction needs to be considered. 
% remain constant in intensity and experience free-space diffraction in the medium. 

Considering the continuous wave limit, the propagation equations for the quantum field $\hat{a}$ reads
% \begin{subequations}
 \begin{align}
 \label{propagation-eq}
  \left(\frac{\partial}{\partial z} - \frac{i}{2k_{p}}\nabla^2_{\perp} \right)\hat{a}(\bm{r}_{\perp},z) &= \frac{ig\mathcal{N}}{c}[\hat{\sigma}^{(1)}_{23}(\bm{r}) +\hat{\sigma}^{(1)}_{14}(\bm{r}) ]\,,
 \end{align}
% \end{subequations}
here $\nabla^2_{\perp}$ introduces the paraxial diffraction which would de-focus the probe in spatial domain. $\hat{\sigma}^{(1)}_{23}(\bm{r})$ and $\hat{\sigma}^{(1)}_{14}(\bm{r})$ which denote the atomic coherence are given by 
\begin{subequations}
 \begin{align}
  \hat{\sigma}^{(1)}_{23}(\bm{r}) &= g\chi_{l1}(\bm{r})\hat{a}(\bm{r}) + g\chi_{n1}(\bm{r})\hat{a}^{\dagger}(\bm{r}) + \hat{F}_{1}(\bm{r})\,,\\[2ex]
  \hat{\sigma}^{(1)}_{14}(\bm{r}) &= g\chi_{l2}(\bm{r})\hat{a}(\bm{r}) + g\chi_{n2}(\bm{r})\hat{a}^{\dagger}(\bm{r}) + \hat{F}_{2}(\bm{r})\,,
 \end{align}
\end{subequations}
$\chi_{j}(\bm{r})$ with $i\in\{l1,l2,n1,n2\}$ describes respectively the linear and nonlinear susceptibilities of the atoms whose exact expressions are usually complicated depending on the laser parameters and are given in the Supplement Material (SM). 
Furthermore, all $\chi_{j}(\bm{r})$ are now spatial-dependent, not only on the transverse coordinates $\bm{r}_{\perp}$ but also the propagation direction $z$ since we have considered a spatial-distributed control field $\Omega_{c1}(\bm{r})$. And $\hat{F}_{1}(\bm{r})$ and $\hat{F}_{2}(\bm{r})$ are the corresponding quantum noise terms respectively. In the following, we will consider the case when the laser parameters are tuned such that the linear and nonlinear absorptions are negligible, thus we may first drop the quantum noise terms. Note here we have assumed the phase-matching condition in the $z$ direction $k_{c1z}+k_{c2z}=2k_{pz}$ with $k_{jz}$ being the propagation wavevector of the field $j$~($j\in\{c1,c2,p\}$). Then the wave equation is modified to 
 \begin{align}
 \label{propagation-eq}
  \left(\frac{\partial}{\partial\zeta} + \frac{i}{2}\nabla^2_{\bm{\xi}} \right)\hat{a}(\bm{\xi},\zeta) &= i\chi_{l}(\bm{\xi},\zeta)\hat{a}(\bm{\xi},\zeta) + i\chi_{n}(\bm{\xi},\zeta)\hat{a}^{\dagger}(\bm{\xi},\zeta)\,,
 \end{align}
where we have rescaled the spatial coordinates as $\bm{\xi}=\bm{r}_{\perp}/S_{\perp}$ and $\zeta=z/S_{z}$ with $S_{z}=k_{p}S^{2}_{\perp}$, and $\chi_{j}=g^{2}\mathcal{N}S_{z}(\chi_{j1}+\chi_{j2})/c$ with $j\in\{l,n\}$ are real functions. Apparently, it is impossible to obtain an analytical solution for the wave equation~(\ref{propagation-eq}), even a numerical calculation turns out to be already very challenging due to the spatial-dependent susceptibilities. Nevertheless, we have managed to numerically solve Eq.~(\ref{propagation-eq}) based on the decomposition of the quantum field $\hat{a}(\bm{\xi},\zeta)$ into a complete set of orthogonal spatial modes
\begin{align}
 \hat{a}(\bm{\xi},\zeta)=\sum_{j}\hat{a}_{j}(\zeta)u_{j}(\bm{\xi},\zeta)
\end{align}
where $a_{j}(\zeta)$ is the annihilation operator at propagation distance $\zeta$ for the $j$th spatial mode $u_{j}(\bm{\xi},\zeta)$ satisfying $\iint^{\infty}_{-\infty}d\bm{\xi}u^{*}_{j}(\bm{\xi},\zeta)u_{l}(\bm{\xi},\zeta)=\delta_{jl}$ and $\sum_{j}u^{*}_{j}(\bm{\xi},\zeta)u_{j}(\bm{\xi}^{'},\zeta)=\delta(\bm{\xi}-\bm{\xi}^{'})$. In principle, $\{u_{j}(\bm{\xi},\zeta)\}$ can be any complete set of functions satisfying the orthogonal relations. For the sake of simplicity, here $u_{j}(\bm{\xi},\zeta)$ is chosen as the eigenfunctions of the paraxial wave equation, i.e., $(\partial_{\zeta}-i\nabla^{2}_{\bm{\xi}}/2)u_{j}(\bm{\xi},\zeta)=0$ such that the diffraction term can be canceled out, leading to 
\begin{align}
\label{mode-eq}
 \frac{d\hat{\bm{A}}(\zeta)}{d\zeta}=i\bm{M}(\zeta)\hat{\bm{A}}(\zeta)
\end{align}
where $\hat{\bm{A}}(\zeta)=\{\hat{a}_{0}(\zeta),\hat{a}_{1}(\zeta),\cdots,\hat{a}_{N-1},
\hat{a}^{\dagger}_{0}(\zeta),\hat{a}^{\dagger}_{1}(\zeta),\cdots,\\\hat{a}^{\dagger}_{N-1}\}$ with $N$ being the number of modes needing to be considered. And
\begin{align}
 \bm{M}(\zeta)=\begin{bmatrix}
            \bm{C}(\zeta) & \bm{D}(\zeta)\\[2ex]
            -\bm{D}^{*}(\zeta) & -\bm{C}^{*}(\zeta)
           \end{bmatrix}
\end{align}
is the propagation matrix determining the output quantum field. $\bm{C}(\zeta)$ and $\bm{D}(\zeta)$ are defined as follows
\begin{subequations}
\begin{align*}
c_{pq}(\zeta) =& \iint^{\infty}_{-\infty}d\bm{\xi} u^{*}_{p}(\bm{\xi},\zeta)\chi_{l}(\bm{\xi},\zeta)u_{q}(\bm{\xi},\zeta)\\[2ex]
d_{pq}(\zeta) =& \iint^{\infty}_{-\infty}d\bm{\xi} u^{*}_{p}(\bm{\xi},\zeta)\chi_{n}(\bm{\xi},\zeta)u^{*}_{q}(\bm{\xi},\zeta)
\end{align*}
\end{subequations}
with $p,q\in\{0,1,\cdots,N-1\}$, and we have $c_{pq}=c^{*}_{qp}$ and $d_{pq}=d_{qp}$ for real $\chi_{l}$ and $\chi_{n}$. Then the formal solution of Eq.~(\ref{mode-eq}) can be given as 
\begin{align}
\hat{\bm{A}}(\zeta)=e^{i\int^{\zeta}_{0}d\zeta^{'}\bm{M}(\zeta^{'})}\hat{\bm{A}}(0)\,.
\end{align}
Furthermore, the effective Hamiltonian for Eq.~(\ref{mode-eq}) can be written as
\begin{align}
\label{quantum-eff-hamiltonian}
 \hat{H}_{\text{eff}}(\zeta)=\frac{1}{2}\sum^{N}_{p,q=1}c_{pq}(\zeta)\hat{a}^{\dagger}_{p}\hat{a}_{q}+d_{pq}(\zeta)\hat{a}^{\dagger}_{p}\hat{a}^{\dagger}_{q} + h.c.
\end{align}
$\hat{H}_{\text{eff}}$ clearly shows that each spatial mode is coupled to all modes including itself, where the first terms indicate photon redistributions between different modes due to the excitation of higher modes and account for classical physics like the optically induced waveguide effects, and the second terms introduce the quantum effects, i.e., quadrature squeezing for all modes and account for all the quantum effects that will be illustrated below including beam focusing, enhanced reduction of local intensity fluctuation, and reduction of beam width uncertainty. 
% In the following, We may term as intra-mode and inter-mode squeezing for $p=q$ and $p\neq q$ respectively. As will be shown below, the intra-mode squeezing leads to the beam defocusing which will be canceled out by the inter-mode squeezing that gives rise to the effect of beam focusing.
% 
% 
\begin{figure}[t!]
 \centering
 \includegraphics[width=0.48\textwidth]{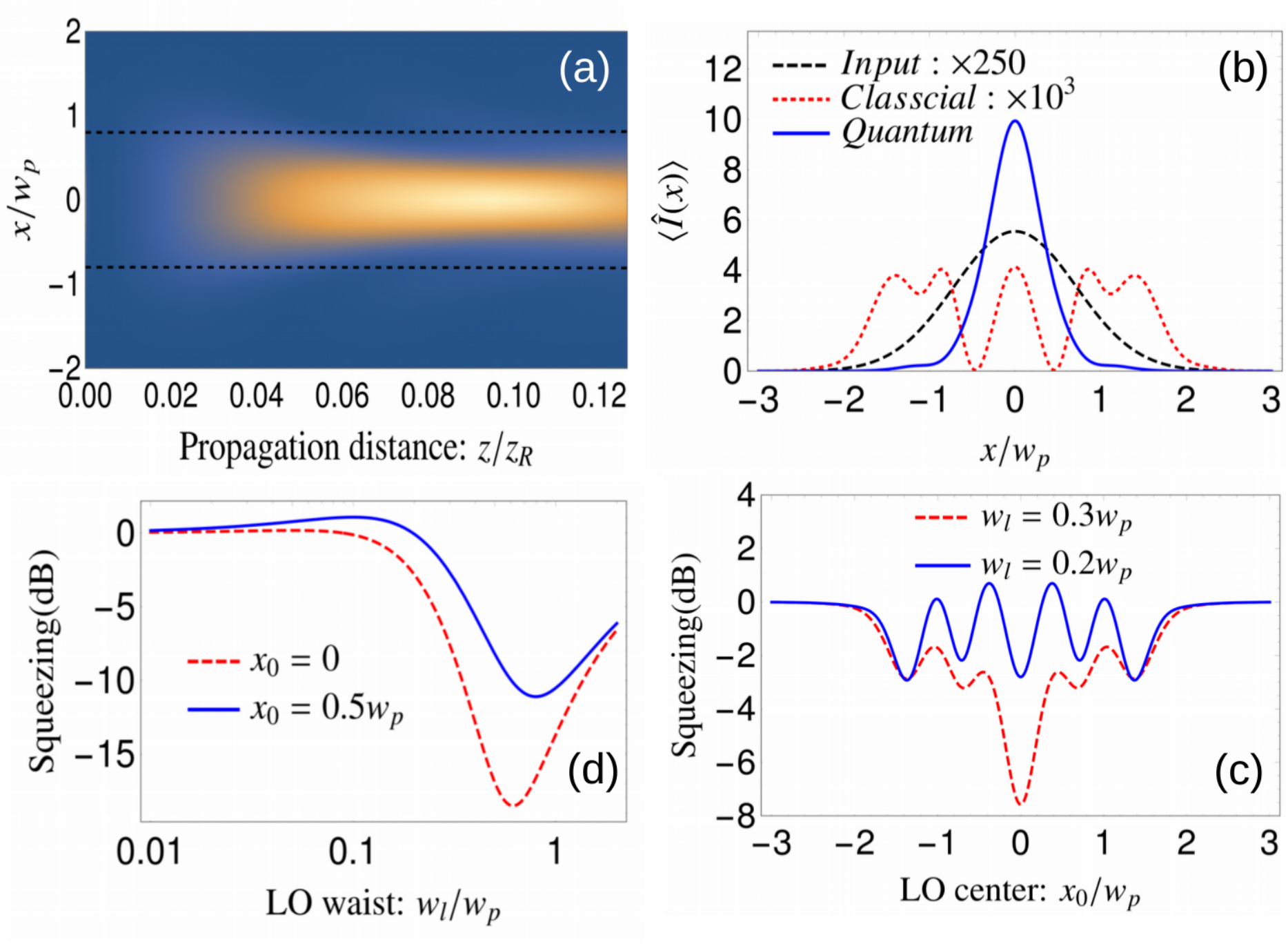} 
 \caption{(Color online) The effect of beam focusing as a result of MSM squeezing. In (a) we show the full propagation dynamics of the quantum probe intensity $\langle\hat{a}^{\dagger}\hat{a}\rangle$ which are gradually focused and enhanced. The output probe is plotted in (b) for both the classical and quantum cases in order to exclude the classical effects. The unique feature of MSM squeezing is illustrated in both (c) and (d) where the squeezing of $\hat{a}_{p}$ is obtained versus the width and central position of the LO. Parameters are: the atomic density $n=3.0\times10^{17}~\text{m}^{-3}, L= 1.0~\text{cm}, S_{\perp}=w_{p}=0.1~\text{cm}, \Gamma_{32}=2\pi\times6.0~\text{MHz},\Omega^{(0)}_{c1}=3\Gamma_{32},\Omega_{c2}=10\Gamma_{32},\Delta_{c1}=41.4\Gamma_{32},\Delta_{c2}=-50\Gamma_{32}, \alpha_{0}= 0.2$.}
 \label{field-dynamics}
\end{figure}
Practically, it is convenient to choose the control field $\Omega_{c1}$ to be a fundamental Hermite-Gaussian beam, i.e.,
\begin{align*}
\Omega_{c1}(x,z)= \frac{w_{c}\Omega^{(0)}_{c1}}{w_{c}(z)}e^{-\bm{r}^{2}_{\perp}/(2w^{2}_{c}(z))}e^{-ik_{c}\bm{r}^{2}_{\perp}/(4R^{2}_{c}(z))}e^{i\phi(z)} 
\end{align*}
with the propagation dependent width $w_{c}(z)=w_{c}\sqrt{1+(z/z_{C})^2}$, radius of curvature $R_{c}(z)=(z^2+z^{2}_{c})/z$ and the Gouy phase $\phi(z)=\arctan(z/z_{C})$, here $z_{C}$ is the associated Rayleigh length. The choice of a Gaussian control beam will lead to a fully symmetric spatial distribution in the transverse plane for an input Gaussian probe, sufficing us to only consider the 2D~($x,z$) propagation dynamics. In order to realize beam focusing, here we chose the spatial size of $\Omega_{c1}(\bm{r}_{\perp},z)$ as $w_{c}=0.8w_{p}$ which is smaller than that of the probe $\hat{a}$. Furthermore, the input Gaussian probe is assumed to be in a coherent state, i.e., $\langle\hat{a}_{j}\rangle=\alpha_{0}\delta_{j0}$ with $\alpha_{0}$ being the amplitude. We have also chosen the Hermite-Gaussian mode basis for $u_{j}(\bm{\xi},\zeta)$ and set $N=40$ at which the numerical solutions have already converged. The numerical results are shown in Fig.~(\ref{field-dynamics}). Fig~\ref{field-dynamics}(a) plots the field intensity $\langle\hat{I}(x,z)\rangle=\sum_{jl}\langle\hat{a}^{\dagger}_{j}(z)\hat{a}_{l}(z)\rangle u^{*}_{j}(x,z)u_{l}(x,z)$ for a short distance $L=0.126z_{R}$ with $z_{R}=7.9\text{ cm}$ being the Rayleigh length for the quantum probe, it can be seen that the laser beam is gradually focused and amplified during propagation. To be more clearly, we have plotted the input and output probe in Fig.~\ref{field-dynamics}(b) where the output probe have been focused to a spatial width $\sqrt{\langle\hat{W}(L)\rangle}\simeq0.55w_{p}$, here the width is defined in terms of the spatial variance of the intensity distribution as follows according to Ref.~\cite{Chille2015OE}
\begin{align}
\label{width-definition}
 \hat{W}(\zeta) =&\frac{1}{\langle\int^{\infty}_{-\infty}\hat{I}(\xi,\zeta) d\xi\rangle } \int^{\infty}_{-\infty}f(\xi)\hat{I}(\xi,\zeta)d\xi\nonumber\\[2ex]
				  =&\frac{1}{\sum_{j}\langle\hat{a}^{\dagger}_{j}\hat{a}_{j}\rangle } \sum_{jl}\hat{a}^{\dagger}_{j}\hat{a}_{l}\int^{\infty}_{-\infty}f(\xi)u^{*}_{j}(\xi,\zeta)u_{l}(\xi,\zeta)d\xi
\end{align}
where $f(\xi)$ is a measure function which here is chosen as $f(\xi)=2\xi^{2}$ such that $\sqrt{\langle\hat{W}(0)\rangle}=w_{p}$. It should be noted $\hat{W}(\zeta)$ has the dimension of an area but not of a length under this choice. In principle one can also consider different appropriate measures $f(\xi)$~\cite{Chille2015OE}.

As mentioned above, the control field $\Omega_{c1}$ is taken to be smaller than the probe in beam size, meaning that parts of the probe lie outside the optical waveguide induced by $\Omega_{c1}$. Thus in the classical picture the probe should not be focused. In order to exclude the possibility that the beam focusing is indeed not induced by classical waveguide effects, we have also calculate the classical field propagation dynamics as shown by the red dashed line in Fig.~\ref{field-dynamics}(b). Evidently, the output field spreads and is totally distorted when only taking the classical dynamics into account, due to the excitations of higher-order spatial modes. However, the situation is essentially different in the quantum regime where all excitations are accompanied by quadrature squeezing. 
As demonstrated by the blue solid line, the quantum probe is considerably narrowed in width. Additionally, the probe is significantly amplified as illustrated by the magnificent factors for the input and classical output probes.
 
The beam focusing can be understood directly in terms of local MSM squeezing in the transverse plane as further shown in Fig.~\ref{field-dynamics}(c) and (d) where we plot the squeezing as a function of the waist and central position of the local oscillator~(LO) with Gaussian distribution $f_{\text{LO}}(x_{0},x)\propto e^{(x-x_{0})^2/(2w^{2}_{l})}$ respectively. Here the degree of squeezing $S$ is defined as 
\begin{align}
 S(x_{0},\zeta)=10\log_{10}\frac{\langle\Delta\hat{P}^2(x_{0},\zeta)\rangle}{\langle\Delta\hat{P}^2(x_{0},0)\rangle}
\end{align}
with $\hat{P}(x_{0},\zeta) \propto i\int^{\infty}_{-\infty}dx[\hat{a}(x,\zeta)f_{\text{LO}}(x_{0},x)e^{-i\theta}-h.c.]$.
As can be seen from Fig.~\ref{field-dynamics}(c), for $w_{l}=0.3w_{p}$ the squeezing for the output probe reaches the maximum at the central area, and then oscillates and eventually decreases gradually to 0, meaning stronger amplification in the probe center as compared to the two wings, and consequently the beam focusing. It should be emphasized here the beam focusing is not possible for single-mode squeezing where the beam width should remain as a constant as suggested by Eq.~(\ref{width-definition}). Furthermore, the spatial oscillation in squeezing can not be observed in single-mode squeezing where squeezing should decreases monotonically as LO shifts away from the probe center, and this is indeed due to the interference between squeezing of different spatial modes. In order to show the MSM nature of squeezing more clearly, we reduce the LO waist to $w_{l}=0.2w_{p}$ where stronger spatial oscillations can be seen. 
% 
% the inhomogeneity of local spatial squeezing across the probe shown in Fig.~\ref{field-dynamics}(d) is straightforward: As we have chosen a Gaussian control field $\Omega_{c1}$ whose intensity is peaked in its center where the FWM process is most efficient, accordingly the probe is maximally squeezed in the center; As the intensity of $\Omega_{c1}$ drops rapidly from the center, the FWM process becomes weaker and eventually vanishes, leading to gradually declined squeezing in the two wings. 
% 
% In order to show the MSM nature of squeezing in a more straightforward way, 
We also calculate the squeezing against the ratio between the spatial size of the LO and that of the probe also shown in Fig.~\ref{field-dynamics}(d). It can be seen that the degree of squeezing reduces rapidly as $w_{l}/w_{p}$. 
% Nevertheless, 0.1~dB squeezing can be still observed at $w_{l}/w_{p}=0.1$, suggesting there are roughly $N\simeq w^{2}_{p}/w^{2}_{l}=100$ squeezed spatial modes. 
Again, depending on the spatial location of the LO, the squeezing degree will be different due to the interference between different modes as we explained above.

% Alternatively, the beam focusing may be also understood as a result of destructive interference between different spatial modes. To demonstrate the interference effects, we divide the intensity into two parts $\hat{I}_{1}(x,z)=\sum_{j}\langle\hat{a}^{\dagger}_{j}(z)\hat{a}_{j}(z)\rangle |u_{j}(x,z)|^{2}$ and $\hat{I}_{2}(x,z)=\sum_{j\neq l}\langle\hat{a}^{\dagger}_{j}(z)\hat{a}_{l}(z)\rangle u^{*}_{j}(x,z)u_{l}(x,z)$ and plot them in Fig.~\ref{field-dynamics}(c). It can be seen that $\langle \hat{I}_{1}\rangle$ is considerably broadened which only accounts for the excitations of different modes, however, $\langle\hat{I}_{2}\rangle$ responsible for the inter-mode contributions is negative in the two wings and has a strong peak in the center due to destructive interference. The sum of $\langle\hat{I}_{1}\rangle$ and $\langle\hat{I}_{2}\rangle$ gives rise to the noticeable narrowing of the probe, namely, beam focusing. 
%
% \begin{figure}[t!]
%  \centering
%  \includegraphics[width=0.48\textwidth]{PhotonNumbersAndPhases.pdf} 
%  \caption{(Color online) Conjugate FWM process where the two control fields $\Omega_{c1}$ and $\Omega_{c2}$ can be considered as classical, the probe and signal fields which are denoted by $\hat{a}^{\dagger}$ and $\hat{a}$ respectively are very weak and should be considered as quantum.}
%  \label{PhotonNumbersAndPhases}
% \end{figure}
% 
\begin{figure}[t!]
 \centering
 \includegraphics[width=0.48\textwidth]{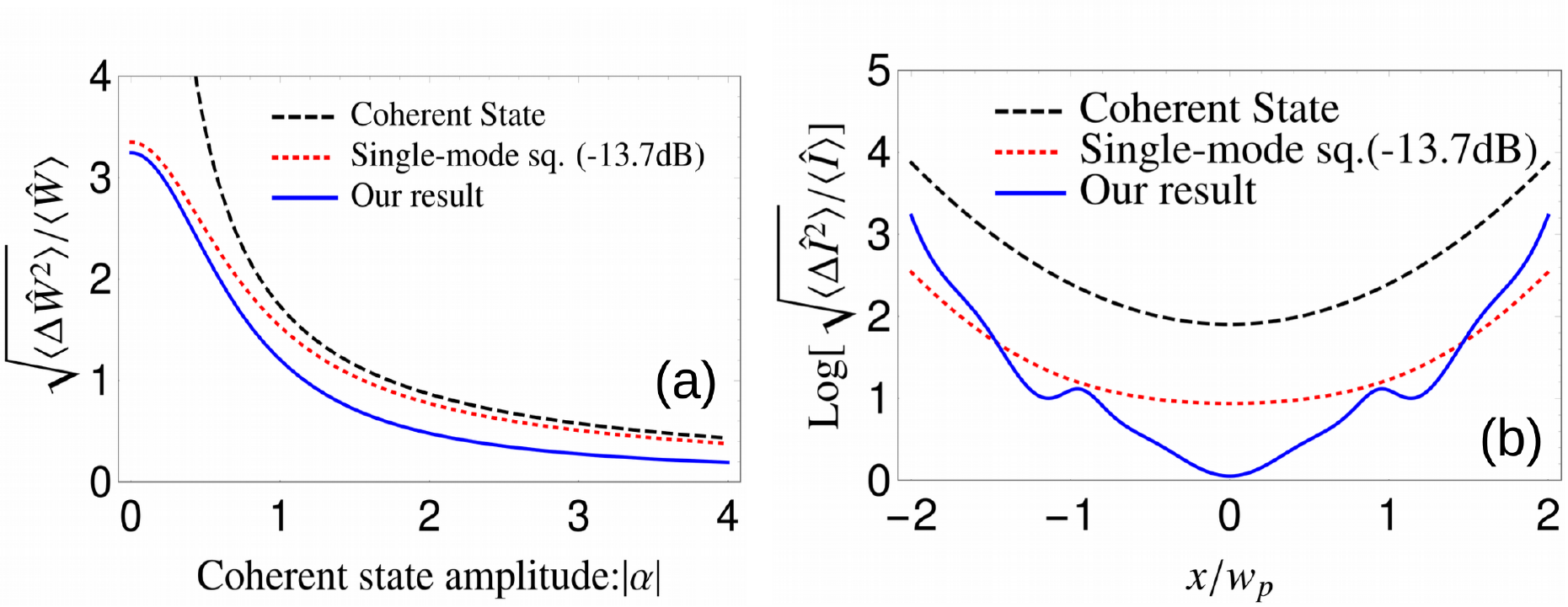} 
 \caption{(Color online) The width uncertainty (a) and local intensity fluctuations (b) versus the amplitude of the coherent incident probe. Parameters are the same as in Fig.~\ref{field-dynamics} except for $\alpha_{0}=0.2e^{i\pi/2}$.}
 \label{fluctuation-reduction}
\end{figure}

The MSM nature of squeezing in the system does not only result in beam focusing, but also remarkable reduction of beam width uncertainty and local intensity fluctuations as depicted in Fig.~\ref{fluctuation-reduction}. In Fig.~\ref{fluctuation-reduction}(a), we calculate the relative reduction of beam width uncertainty $\sqrt{\langle\Delta\hat{W}^{2}\rangle}/\langle\hat{W}\rangle$ for the output probe as a function of the amplitude of the coherent incident probe $|\alpha|$. As a comparison, we also compute the width uncertainty for a single fundamental Hermitian-Gaussian mode which is in either coherent or squeezed state. The squeezed coherent state is chosen to be squeezed in amplitude which results in maximal reduction of width uncertainty~(see SM). As compared to the case of single-mode squeezing of -13.7 dB which is the strongest degree of squeezing that can be obtained in the case of $w_{l}=w_{p}$ in our results as sown Fig.~\ref{field-dynamics}(d), the relative reduction is always more pronounced in the case of MSM squeezing as shown by the blue solid line in Fig.~\ref{fluctuation-reduction}(a). 
In particular, for very weak incident probe at $|\alpha|=1$, the relative width uncertainty is significantly reduced to $47.9\%$ which is smaller than $77.4\%$ or $86.6\%$ for single-mode squeezed or coherent states respectively. 
One should be noted that the absolute reduction of width uncertainty in our result should be approximately doubled since the width of the output probe is reduced to $0.55w_{p}$.
The underlying physics is that the MSM squeezing leads to a much stronger suppression of local intensity fluctuation in the central area of the probe which determines the width uncertainty, in contrast to the single-mode case as plotted in Fig.~\ref{fluctuation-reduction}(b), consequently a remarkable reduction in the width uncertainty which is a weighted summation of the local intensity fluctuation according to Eq.~(\ref{width-definition}) can be obtained.

% We further calculate the effect of beam focusing and reduction of width uncertainty as a function of the waist of the incident control field $\Omega_{c1}$ as shown in Fig.~\ref{field-dynamics}. As can be seen, the probe experiences defocussing for small $w_{c}$ due to the strong diffraction of $\Omega_{c1}$ in the medium. At the meantime, there is little reduction of width uncertainty because of the weak MSM squeezing. When $w_{c}$ becomes larger, the probe is focused with strong reduction of width uncertainty, and then approaches to the extremism $w_{p}$ at $w_{c}$.  

{\it Discussions and Conclusions.} Further enhancement on the beam focusing and reduction of width uncertainty can be expected by tuning the laser parameters including beam width, detuning and intensities of the control beam $\Omega_{c1}(x,z)$, as well as the atomic density. For example, the probe beam may be further tightly focused by employing a control $\Omega_{c1}(x,z)$ with much smaller spatial size, provided that the stronger diffraction of the small-sized $\Omega_{c1}(x,z)$ can be reduced. This would require to use atomic sample of short length, which in turn would require higher atomic density to get accountable amount of MSM squeezing which ensures stronger beam focusing and reduction of width uncertainty. However, in general, a reliable prediction will be very difficult to make, considering that a large number of spatial modes are involved and it is in general a many-body problem with ``time-dependent" interaction between them as indicated by Eq.~(\ref{quantum-eff-hamiltonian}).

In our model, we have restricted ourself in the paraxial regime for proof-of-principle demonstration of simultaneous beam focusing and reduction of width uncertainty. In general, extension to the non-paraxial regime should be possible. However, the numerical calculation will be extremely complicated. As we have mentioned earlier, even in the paraxial regime the numerical simulations become already very challenging.

We have demonstrated the unexpected and surprising effects of simultaneous beam focusing and remarkable reduction of width uncertainty via MSM squeezing in a conjugate FWM process. The beam focusing is achieved due to the MSM squeezing which leads to inhomogeneous spatial squeezing in the transverse plane. Furthermore, the considerable reduction of width uncertainty is due to the localized spatial squeezing which reduce significantly the local quantum fluctuations. Our results can be very useful in quantum metrology and precision measurements such as detection of very small displacement of particles in biological system where only weak quantum light is allowed.

L. Zhang is grateful for the helpful discussions with Tao Peng, Zhenhuan Yi. We acknowledge the support of Office of Naval Research Grant No. N00014-16-1-3054 and Robert A. Welch Foundation Award No. A1261.

\bibliographystyle{apsrev4-1}
\bibliography{referencesbase}

\end{document}